\documentclass[10pt]{article}

\usepackage{amsmath}
\usepackage{amssymb}

\usepackage{graphicx}

\usepackage{cite}

\usepackage{color} 

\usepackage{fancyhdr}

\usepackage[labelfont=bf,labelsep=period,justification=raggedright]{caption}

\makeatletter
\renewcommand{\@biblabel}[1]{\quad#1.}
\makeatother

\date{}

\begin{document}

\thispagestyle{fancy}
\rhead{Young GF, Scardovi L, Cavagna A, Giardina I, Leonard NE (2013) Starling Flock Networks Manage Uncertainty in Consensus at Low Cost. PLoS Comput Biol 9(1): e1002894}

\begin{flushleft}
{\Large
\textbf{Starling flock networks manage uncertainty in consensus at low cost}
}
\\
George F. Young$^{1}$, 
Luca Scardovi$^{2}$,
Andrea Cavagna$^{3}$, 
Irene Giardina$^{3}$,
Naomi E. Leonard$^{1,\ast}$
\\
\bf{1} Department of Mechanical and Aerospace Engineering, Princeton University, Princeton, NJ, USA
\\
\bf{2} Department of Electrical and Computer Engineering, University of Toronto, Toronto, ON, Canada
\\
\bf{3} Istituto dei Sistemi Complessi, Consiglio Nazionale delle Ricerche and Dipartimento di Fisica, Universit\'{a} Sapienza, Rome, Italy
\\
$\ast$ E-mail: naomi@princeton.edu
\end{flushleft}

\section*{Abstract}
Flocks of starlings exhibit a remarkable ability to maintain cohesion as a group in highly uncertain environments and with limited, noisy information. Recent work demonstrated that individual starlings within large flocks respond to a fixed number of nearest neighbors, but until now it was not understood why this number is seven. We analyze robustness to uncertainty of consensus in empirical data from multiple starling flocks and show that the flock interaction networks with six or seven neighbors optimize the trade-off between group cohesion and individual effort. We can distinguish these numbers of neighbors from fewer or greater numbers using our systems-theoretic approach to measuring robustness of interaction networks as a function of the network structure, i.e., who is sensing whom. The metric quantifies the disagreement within the network due to disturbances and noise during consensus behavior and can be evaluated over a parameterized family of hypothesized sensing strategies (here the parameter is number of neighbors). We use this approach to further show that for the range of flocks studied the optimal number of neighbors does not depend on the number of birds within a flock; rather, it depends on the shape, notably the thickness, of the flock. The results suggest that robustness to uncertainty may have been a factor in the evolution of flocking for starlings. More generally, our results elucidate the role of the interaction network on uncertainty management in collective behavior, and motivate the application of our approach to other biological networks.

\section*{Author Summary}
Starling flocks move in beautiful ways that both captivate and intrigue the observer. Previous work has shown that starlings pay attention to their seven closest neighbors, but until now it was not understood why this number is seven. Our paper explains the mystery: when uncertainty in sensing is present, interacting with six or seven neighbors optimizes the balance between group cohesiveness and individual effort.  To prove this result we develop a new systems-theoretic approach for understanding noisy consensus dynamics. The approach allows the evaluation of robustness over a family of hypothesized sensing strategies using observations of the spatial positions of birds within the flock. We apply this approach to experimental data from wild starling flocks, and find that six or seven neighbors yield maximal robustness.  The implication that robustness of cohesion may have been a factor in the evolution of flocking has significant consequences for evolutionary biology. In addition, the results and the versatility of the approach have implications for uncertainty management in social and technological networks.

\section*{Introduction}
Flocks of birds and schools of fish exhibit striking and robust collective behaviors despite the challenging environments in which they live~\cite{Ballerini08, Bialek12, Partridge81, Partridge80, Hemelrijk11, Cucker08, Conradt03, Couzin05, Sumpter08}. These collective behaviors are believed to emerge from simple, local interactions among the individuals~\cite{Okubo86, Deneubourg89, Grunbaum98, Parrish99, Couzin03}. Significantly, such groups are able to maintain cohesion and coherence even when every individual is subject to uncertain information about the behavior of its neighbors (those  in the group that it can sense) as well as disturbances from the environment. However, it is not well understood if and how this robustness to uncertainty depends on the structure of the interaction network of individuals, that is, on who is sensing and responding to whom. 

Recent analysis of position~\cite{Ballerini08} and velocity~\cite{Bialek12} correlations in empirical data collected for large flocks of starlings (\emph{Sturnus vulgaris}) has shown that each bird responds to a fixed number, seven, of its nearest neighbors. This work suggests that following a topological interaction rule (i.e.\ interacting with a fixed number of neighbors) provides important robustness benefits for group cohesion compared to a metric rule (i.e.\ interacting with neighbors within a fixed distance)~\cite{Ballerini08}. In addition, work is underway on techniques that can reveal in greater detail the precise nature of the inter-individual interactions~\cite{Eriksson10, Bode12}. However, these analyses do not yield an explanation for why the starlings interact with seven neighbors, rather than some other number of neighbors.

Here, we address the question of what is the connection between the number of neighbors used by each bird for social information and the robustness of the flock as a whole. We evaluate robustness for starling flocks using three-dimensional positions of birds studied in \cite{Ballerini08,Bialek12} and a metric that quantifies, as a function of the interaction network, the ability of a group to achieve consensus in the presence of uncertainty. Our metric derives from the common assumption that each individual bird carries out a consensus-type behavior in response to its neighbors, and that a linear dynamical system  describes how the interaction network allows the group to reach consensus.  We introduce noise into the linear dynamical model and quantify robustness to uncertainty by the resulting disagreement within the group.

Our systems-theoretic approach makes it possible to evaluate robustness to uncertainty over a parameterized family of hypothesized individual sensing strategies given observations of the group.   For the starling flocks we evaluate the set of  strategies corresponding to each individual sensing and responding to a fixed number of closest neighbors.  Since the interaction structure of each starling flock network is determined by the measured spatial distribution of the birds and the strategy that each bird uses to determine which neighbors it senses, we can apply our metric to the starling flock data to distinguish which strategy (i.e., which number of neighbors), among a parameterized family of strategies (i.e., the family parameterized by number of neighbors), minimizes the influence of  uncertainty on how close the birds come to consensus.

Assuming that every bird in a flock responds to a fixed number of neighbors (\textit{m}) and that each interaction poses some cost in effort to the bird, we compute the per-neighbor contribution to robustness as a function of \textit{m}.  The interaction cost, accounted for by the per-neighbor calculation, is associated with the understanding that achieving consensus is not the only behavior undertaken by the birds: in addition to remaining with the flock, each bird must watch for and avoid predators,  seek food or a roosting site, etc. Thus, the flock must be responsive to external signals in addition to remaining cohesive, and this requires that each individual use as little effort as possible for maintaining cohesion.  We show that across all flocks in the data set, interaction networks with six or seven neighbors maximize the per-neighbor contribution to robustness. 

By analyzing variations between different flocks, we show further that for the range of flocks observed the optimal number of neighbors (\textit{m}*) does not depend on the size of a flock (\textit{N}). Instead, both the optimal number of neighbors and the peak value of robustness per neighbor depend on the shape (in particular the thickness) of the flock.

\section*{Methods}
Most models of flocking are based on consensus behavior~\cite{Hemelrijk11, Cucker08, Couzin05, Vicsek95, Czirok00, Leonard12}.  Accordingly, we define our robustness metric for a flock carrying out linear consensus dynamics~\cite{Moreau05, OlfatiSaber04, Jadbabaie03} on some quantity of interest, such as a direction of travel. It should be noted that the following analysis applies equally to any variable of interest. Each individual maintains a copy of the variable to be agreed upon. In a group of \textit{N} individuals, the $i^\text{th}$ individual then evolves its own variable ($x_i$) over time \textit{t} according to the weighted sum of the differences between its variable and those of its neighbors, according to
\begin{equation} \label{eqn:dynamics}
\frac{\mathrm{d} x_i}{\mathrm{d}t} = \sum_{j \in N_i} a_{i,j} (x_j - x_i) + \xi_i,
\end{equation}
where $N_i$ is the set of neighbors of the $i^\text{th}$ individual, $a_{i,j}$ is the weight given by individual \textit{i} to information from individual \textit{j}, and $\xi_i$ is a source of noise~\cite{Young10}.

Recent work has shown that Eq.~(\ref{eqn:dynamics}) is the minimal model consistent with experimental correlations in natural flocks of birds~\cite{Bialek12}. Even though directions are inherently nonlinear quantities, linear consensus is a good model for consensus on direction of travel when the relative differences in directions are small~\cite{Moreau05}. Representing each individual as a node, and a directed edge with weight $a_{i,j}$ from node \textit{i} to node \textit{j} whenever $a_{i,j}$ is non-zero, we obtain a weighted, directed sensing graph that encodes all information transfer between individuals within the flock and thus defines the interaction network.  The properties of this sensing graph are intimately related to the ability of the group to achieve consensus; for example, consensus can be reached when noise is absent if and only if the graph is connected~\cite{Moreau05, OlfatiSaber04, Jadbabaie03}. 

By defining an \textit{N}-dimensional state vector \textbf{x}, containing the variables $x_i$ from each individual in the network, we can combine Eq.~(\ref{eqn:dynamics}) from each individual into the vector equation
\begin{equation}\label{eqn:matrixdynamics}
\frac{\mathrm{d} {\mathbf x}}{\mathrm{d}t} = -L {\mathbf x} + {\boldsymbol \xi},
\end{equation}
where ${\boldsymbol \xi}$ is a vector containing the individual noise terms $\xi_i$ and \textit{L} is known as the Laplacian matrix of the graph~\cite{OlfatiSaber04, Jadbabaie03, Young10}. The Laplacian matrix  is commonly used to encode a graph:  the $(i,j)^\text{th}$ element of \textit{L} for $i \neq j$ is equal to $-a_{i,j}$, the negative of the weight on the $(i,j)^\text{th}$ edge (or 0 if this edge is not present), and the $i^\text{th}$ diagonal entry of \textit{L} is the out-degree for the $i^\text{th}$ node, $\sum_{j\in N_i} a_{i,j}$, i.e., the sum of the weights on all edges leaving the $i\text{th}$ node. 

In this setting, consensus corresponds to \textbf{x} having every entry equal, i.e., \textbf{x} is a scalar multiple of the vector containing all ones. The set of every possible consensus state is thus a one-dimensional subspace of the \textit{N}-dimensional state space. The disagreement in the system is measured by the minimum distance (in \textit{N}-dimensional space) from the state to this ``consensus'' subspace. When noise is present, the system will conduct a random walk, which may or may not remain close to consensus.   Our robustness metric quantifies how close to consensus this random walk remains as a function of the interaction network encoded by \textit{L}.

We measure the robustness of consensus to noise by the expected steady-state disagreement when every agent has a unit-intensity i.i.d. source of white noise ($\xi_i$ for the $i^\text{th}$ agent)~\cite{Young10, Bamieh12, Xiao07}. We can analyze the disagreement dynamics by defining an ($N-1$)-dimensional vector ${\mathbf y} = Q{\mathbf x}$, where \textit{Q} is a matrix with rows that form an orthonormal basis for the subspace orthogonal to the consensus subspace. This leads to the following dynamics for \textbf{y}:
\begin{equation}\label{eqn:disagree}
\frac{\mathrm{d}{\mathbf y}}{\mathrm{d}t} = -QLQ^T {\mathbf y} + Q{\boldsymbol \xi}.
\end{equation}
We define $\bar{L} = QLQ^T$, which we call the reduced Laplacian matrix. 

The disagreement in the system is the length of \textbf{y}, and we seek to compute the expected value of this length as time goes to infinity, i.e., at steady state. When the graph is connected, the reduced Laplacian will be a stable matrix~\cite{Young10}, and hence \textbf{y} will converge to a stationary distribution. The covariance matrix of this distribution is the solution, $\Sigma$, to the Lyapunov equation 
\begin{equation}\label{eqn:lyapunov}
\bar{L} \Sigma + \Sigma \bar{L}^T = I,
\end{equation}
and hence the steady-state disagreement, or $\mathcal{H}_2$ norm, is given by~\cite{Young10}
\begin{equation}\label{eqn:H2}
\mathcal{H}_2 = \sqrt{\text{Trace}(\Sigma)}.
\end{equation}
Given an interaction network, encoded by \textit{L}, the reduced Laplacian $\bar{L}$ can be computed and the covariance metric $\Sigma$  determined from Eq.~(\ref{eqn:lyapunov}).  The steady-state disagreement is computed from Eq.~(\ref{eqn:H2}) in the case that every agent's response is corrupted by  noise with intensity 1. If the noise has some intensity other than 1, the resulting $\mathcal{H}_2$ norm is simply scaled by the intensity of the noise.  

The metric depends on \textit{N} since it is a distance in \textit{N}-dimensional space. This dependence is removed by dividing by the square root of \textit{N}, to obtain the expected disagreement due to each individual. By inverting this quantity, we obtain an ``$H_2$ nodal robustness'' which is small (large) when individuals contribute a large (small) amount of disagreement. The robustness is zero precisely when the graph is not connected, and the individuals are unable to reach consensus even in the absence of noise~\cite{Young10}. 

Our robustness metric is most suitable to our purposes:  since the metric only depends on the sensing graph, we can evaluate robustness for different sensing strategies (e.g., choice of \textit{m} neighbors), provided we can construct the resulting graphs.

Previous analysis of the observed positions of starlings within large flocks (440 to 2600 birds) has shown that the birds interact with seven nearest neighbors, irrespective of flock density~\cite{Ballerini08, Bialek12}.  Since the starling data were collected during flocking events with no apparent direct targets or threats to the birds~\cite{Ballerini08}, we assume that a primary goal of each bird was to remain with the flock, i.e., to maintain consensus on a direction of flight. In more complicated scenarios, such as goal-oriented behavior, different metrics may be used to evaluate individual performance, such as an individual's average speed in the direction of the goal~\cite{Codling07}. In addition, other robustness measures, such as the $\mathcal{H}_\infty$ norm, may be more relevant if the disturbances in the system are non-random. However, in our scenario, it is most natural to use the $H_2$ nodal robustness to obtain a measure of how well the starling flock networks managed uncertainty: first we re-construct the sensing graph by applying to the three-dimensional positions of birds the strategy in which each bird uses information from its seven closest neighbors, and then we compute the robustness metric for that graph. We can likewise compute and compare the sensing graph and robustness metric corresponding to any interaction strategy by applying it to the same position data; here we focused on the strategies in which each bird uses information from its \textit{m} closest neighbors, and we examined the set with \textit{m} ranging from 1 to 11.  

It is possible that the birds  weight the information from different neighbors differently, for example, depending on their distance or how well they are sensed.   To be conservative and consistent, we consider that each individual uses an unweighted average of the information from its \textit{m} nearest neighbors, so $a_{i,j}$ equals $\frac{1}{m}$ when an edge is present. In fact, preliminary calculations using other plausible weighting schemes suggest that equal weights lead to better robustness and that the weights must vary significantly between neighbors in order to change our results (Fig.~S1). Further, our calculations evaluate robustness at steady-state for fixed sensing graphs; however, the steady-state assumption in our computation is only required to remove transient dependence on initial conditions. Hence for a group already close to consensus but with a time-varying graph, our steady-state calculation reflects the instantaneous performance of the flock.

The cost for an individual starling to sense the behavior of each neighbor comes from sensory and neurological requirements as well as time lost for watching for predators or searching for a roosting site, etc. It is known that birds have a limited and thus costly capability for tracking multiple objects~\cite{Emmerton93}. To account for these costs, which increase with increasing \textit{m}, we evaluated the ``robustness per neighbor'' for each value of \textit{m}, which is computed as the $H_2$ nodal robustness divided by the number of neighbors \textit{m}. This allows us to identify ranges of \textit{m} of increasing (decreasing) return, where the robustness per neighbor increases (decreases) with \textit{m}. We define the optimum \textit{m} for robustness, \textit{m}*, as the value that maximizes the robustness per neighbor.

We computed robustness per neighbor ($H_2$ nodal robustness divided by \textit{m}, number of neighbors sensed by each bird) for data sets from twelve starling flocks: all ten flocks that were studied in \cite{Ballerini08} and two additional flocks that were studied in \cite{Bialek12}.  From each flock there were between 16 and 80 snapshots over time for a total of 394 snapshots. The number of birds in these flocks ranged from 440 to 2600.

\section*{Results}
For small values of \textit{m}, robustness will be zero since the sensing graph will not be connected~\cite{Xue04}. Robustness per neighbor will increase with increasing \textit{m} to non-zero values when \textit{m} is sufficiently large for the graph to be connected. Further, robustness is bounded above by the value for the complete graph (in which every individual can sense every other individual), and so robustness per neighbor can be expected to be a decreasing function of \textit{m} for large values of \textit{m}. Thus, a priori we can expect a peak in robustness per neighbor as a function of \textit{m}.

\begin{figure}[ht]
\begin{center}
\includegraphics[width=\textwidth]{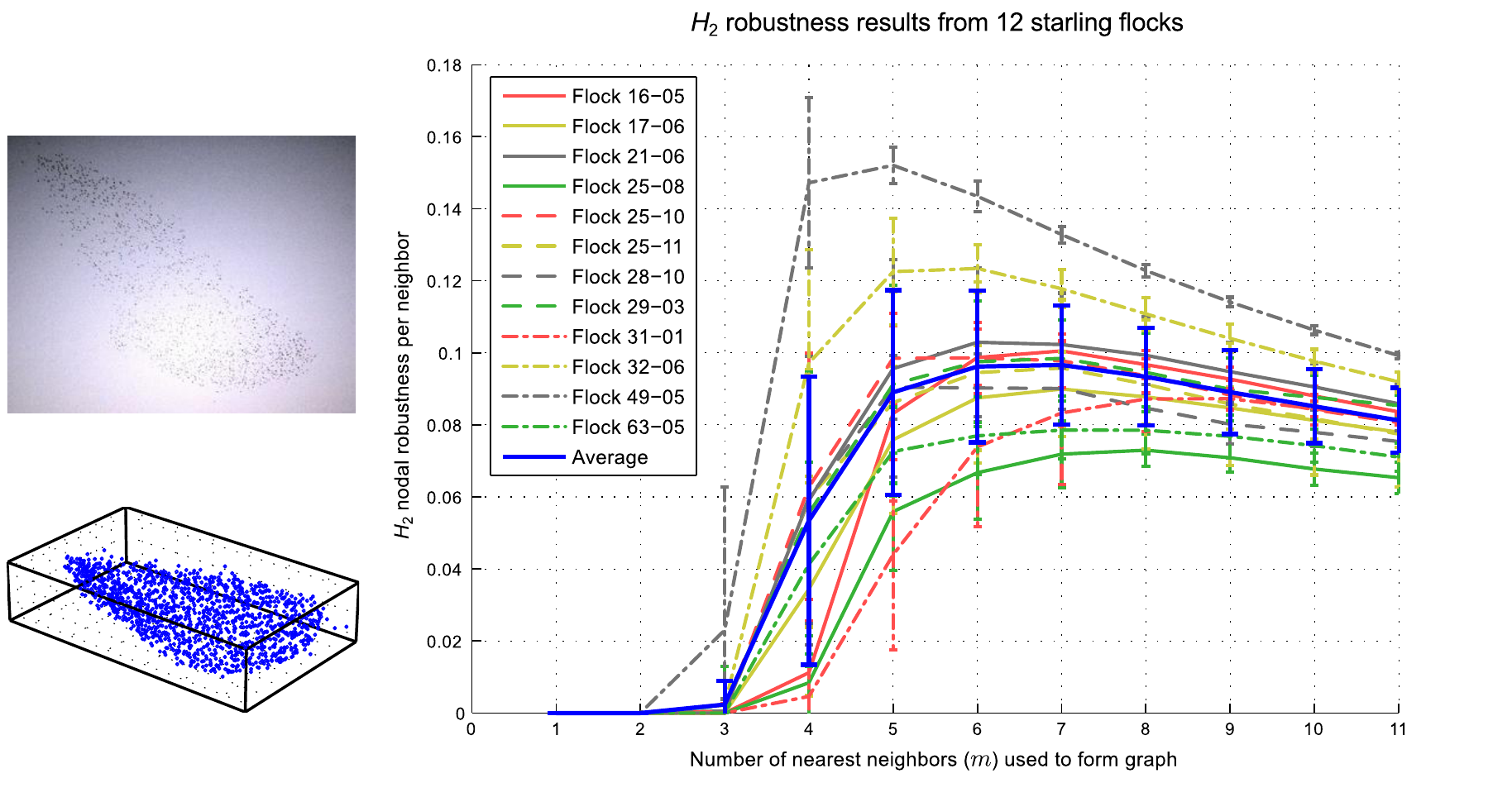}
\end{center}
\caption{
\textbf{$\mathcal{H}_2$ nodal robustness per neighbor as a function of the number of nearest neighbors (\textit{m}) used to form the graph, for twelve separate flocks as well as an overall average.} For each flock the curve shown is the average of all snapshots taken of that flock, with error bars showing the standard deviation. The overall average, shown as the blue curve, is an average of the twelve flocks, with error bars showing the standard deviation. If, instead, an average is taken of every snapshot (394 in total), the resulting curve and standard deviations are almost identical (see Fig.~S2), although the error is greatly reduced (see Fig.~S3). On the left is shown a snapshot of starling flock 25-08 in flight and the corresponding tracked positions, rotated to fit inside a rectangular bounding box.
}
\label{fig:mainresult}
\end{figure}

For all ten flocks studied in \cite{Ballerini08} and two additional flocks studied in \cite{Bialek12}, we computed the robustness per neighbor for each snapshot for $m = 1, \ldots, 11$.   The average robustness per neighbor for each flock is shown in Fig.~\ref{fig:mainresult}, along with the average of the twelve flock averages, as a function of \textit{m}.  In every case, the graphs remained disconnected for \textit{m} equal to 1 and 2, but almost all graphs were connected when \textit{m} was equal to 5. Each flock attained its peak robustness per neighbor value for \textit{m} between 5 and 9 (Fig.~\ref{fig:mainresult}), i.e., at higher values of \textit{m} than were required for connectivity; this demonstrates that our robustness measure is not simply recording the onset of connectivity. The average robustness per neighbor across all flocks reached its peak value at either $m = 7$ or $m = 6$ (Fig.~\ref{fig:mainresult}). Therefore, the observed behavior of the starlings ($m = 7$) from \cite{Ballerini08, Bialek12} places them at a point that maximizes the robustness per neighbor.

We further investigated observed variation in the value of \textit{m}* for different flocks. When the average robustness was computed by averaging every snapshot from every flock, rather than by averaging the flock averages, we obtained almost identical results (Figs.~S2 and S3). This suggested that we could treat all 394 snapshots as independent data points and strengthened the generality of the result in the case in which we treated the flocks as the independent observations. 

\begin{figure}[ht]
\begin{center}
\includegraphics[width=\textwidth]{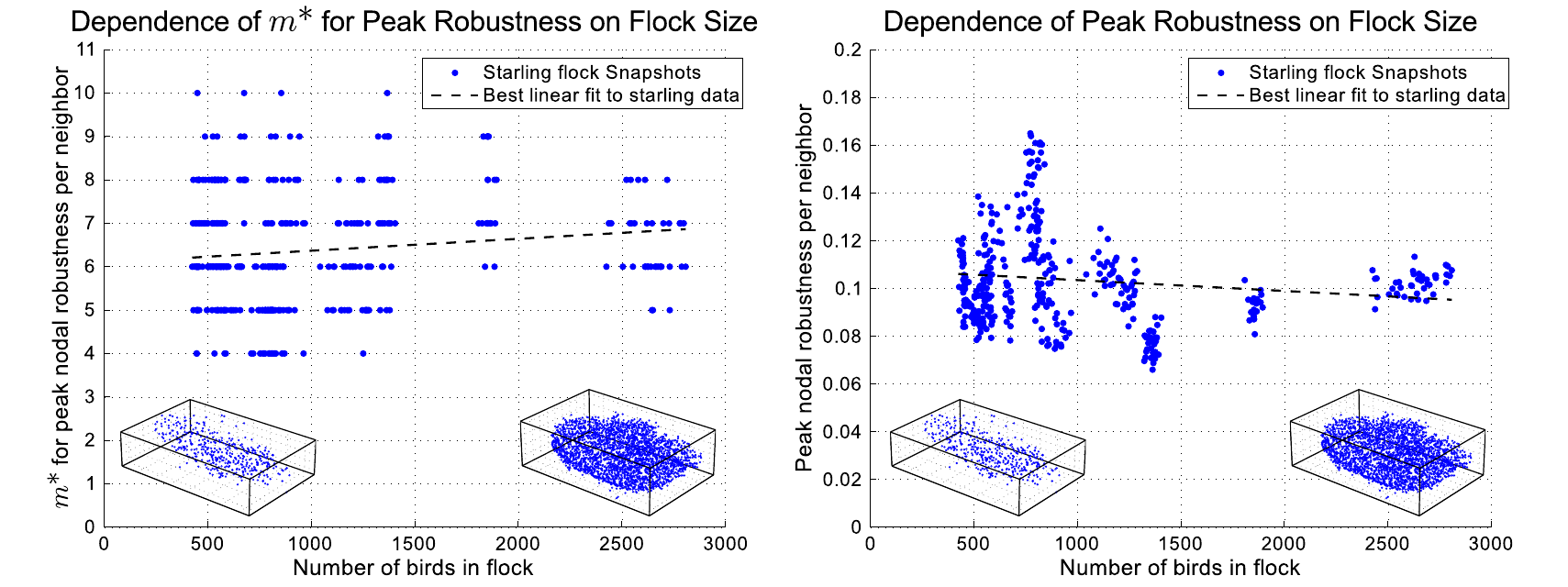}
\end{center}
\caption{
\textbf{Dependence of the optimum number of neighbors (\textit{m}*) and the peak value of robustness per neighbor on the number of birds in the flock (\textit{N}).} Different snapshots from the same flock have different numbers of birds due to occlusions. Results for each snapshot are shown rather than averaged across flocks since we can take each snapshot to be an independent observation (see Fig.~S2). Under each plot are the bird positions (rotated to fit inside a rectangular bounding box) for two snapshots corresponding to the smallest and largest flocks studied.
}
\label{fig:size}
\end{figure}

In a fully random group, the number of neighbors required for connectivity, and hence \textit{m}*, grows weakly with the size of the group (on the order of $\log N$~\cite{Xue04}). However, even when connectivity is attained, noise has a crucial role in determining whether or not  global order can be reached. First, above a certain noise threshold (critical temperature), global order is lost, whether or not the network is connected. Second, even in the low noise phase and on a connected static network, depending on the physical dimension of space and on the topology of the network~\cite{Mermin66, Cassi92} there are cases where order can be reached only if the number of neighbors scales with \textit{N}. Given that our method is static in nature (it does not take into account birds' motion) and that the topology of flocks' network is nontrivial, the dependence of \textit{m}* on \textit{N} may be a concern. However, the variation observed here in \textit{m}* was not a result of varying flock size since neither the value of \textit{m}* nor the peak robustness per neighbor showed a significant dependence on the number of birds in the flock (Fig.~\ref{fig:size}). In both cases, the best linear fit to the data has negligible slope with an $R^2$ value of 0.0178 in the case of \textit{m}* and 0.0230 in the case of peak robustness per neighbor.

\begin{figure}[ht]
\begin{center}
\includegraphics[width=\textwidth]{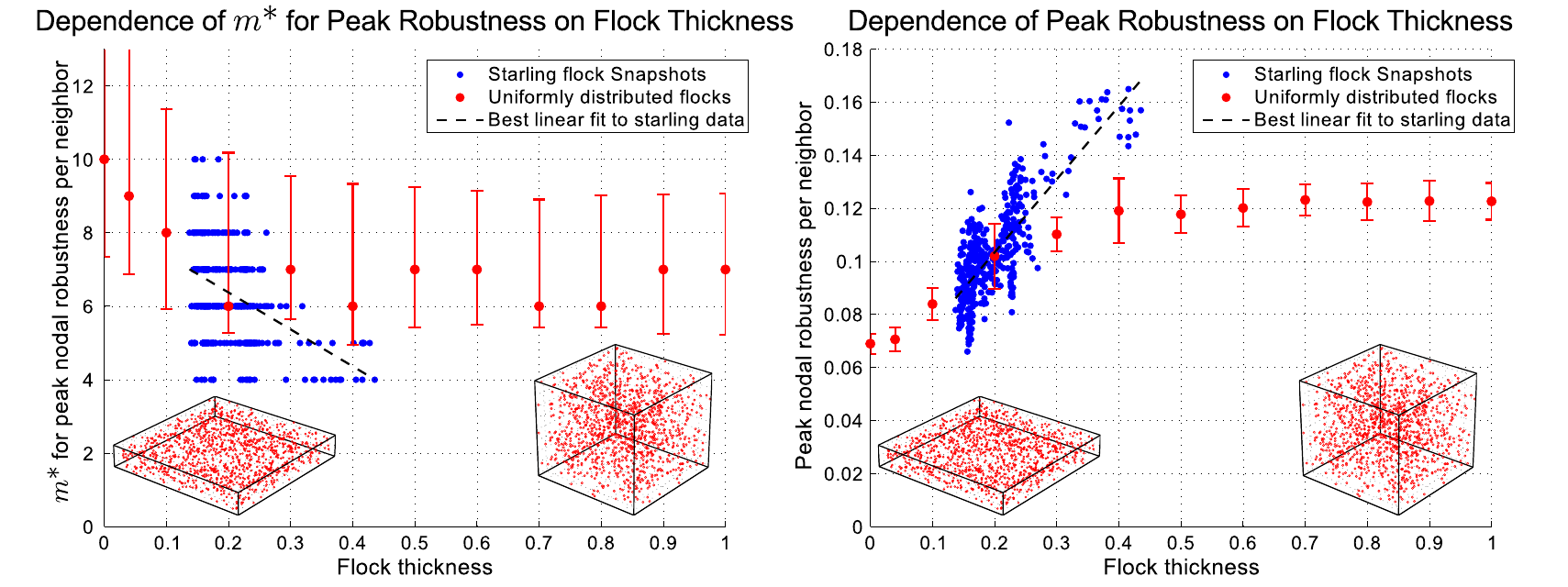}
\end{center}
\caption{
\textbf{Dependence of the optimum number of neighbors (\textit{m}*) and the peak value of robustness per neighbor on the thickness of the flock.} Flock thickness is defined as the ratio of smallest to largest dimension of an ellipsoid having the same principal moments of inertia as the flock.  Results are shown in blue from each snapshot of starling data and in red from flocks randomly generated from a uniform distribution within a rectangular prism. Each data point shown from the random flocks is the average result from generating 100 separate flocks, each containing 1200 individuals. The error bars shown for peak values are the standard deviation, while the error bars for \textit{m}* show the range of values for which the robustness per neighbor is within 90\% of the peak. Under each plot are the positions of two randomly generated flocks, with thicknesses of 0.15 and 0.85.
}
\label{fig:shape}
\end{figure}

Instead we observed a strong dependence of both \textit{m}* and peak robustness per neighbor on flock thickness (Fig.~\ref{fig:shape}). We measured flock thickness as the ratio of smallest to largest dimension of an ellipsoid having the same principal moments of inertia as the flock. Thus a two-dimensional flock has a thickness of 0 while a flock with an equal spread of birds in all directions has a thickness of 1.  We found that the starling flocks had thicknesses between 0.13 and 0.44, with most between 0.13 and 0.27. Across this range, both the variation in \textit{m}* and the average value of \textit{m}* decreased significantly with thickness. The best linear fit to the data displays a negative slope with an $R^2$ value of 0.1816, which is relatively low due to the changing variance. Furthermore, the peak robustness per neighbor increased significantly with thickness (Fig.~\ref{fig:shape}). The best linear fit to the data has a positive slope and an $R^2$ value of 0.6435. No such dependencies were observed with the width of the flock (Figs.~S4 and S5).

To further understand the dependence on thickness, we generated random flocks of varying thickness (with 1200 individuals). For uniformly distributed flocks, \textit{m}* initially decreased with thickness before leveling out, while peak robustness per neighbor showed an increase with thickness according to a sigmoidal shape (Fig.~\ref{fig:shape}). Similar behavior, although with less pronounced variation in \textit{m}*, was observed when using different distributions that have more ordering; with increased ordering the trends show lower values of \textit{m}* and higher values of peak robustness  (Figs.~S6 and S7). The fact that the values of \textit{m}* for the starlings tend to be slightly lower than those of uniformly distributed flocks, while peak robustness tends to be slightly higher, is consistent with the fact that starlings have a more regularly separated distribution than uniformly distributed points.

\section*{Discussion}
Our analysis shows that the size (seven) of each starling's neighborhood~\cite{Ballerini08, Bialek12} optimally trades off gains from robustness with costs associated with sensing and attention; this suggests that robustness to uncertainty may have been a factor in the evolution of flocking. The fact that the same number of neighbors is optimal over a range of flock sizes and densities (as well as, to a certain extent, typical flock thicknesses) suggests that the number of neighbors that a bird interacts with could be an evolved trait.   This is  consistent with the fact that the ability to follow more neighbors requires additional sensory and cognitive apparatus.   A bird that fully utilizes whatever capability it has will contribute most to maximizing the absolute robustness of the group; however, our results provide an explanation for why the evolved capacity of starlings should be limited to seven neighbors.   Further investigation is required to discern whether evolutionary processes could lead to the optimization of efficient robustness at the level of the group.

The trade-off seen here between robustness and sensing cost is not observed for performance metrics related to responsiveness, such as the speed of convergence to consensus (Fig.~S8). Although responsiveness is an important property of group behavior, our results correspond with the previous observation~\cite{Ballerini08} that the primary benefit of the observed interaction rule within starling flocks is to improve robustness. Other aspects of behavior, such as the way in which individuals respond to external signals, may be required for an analysis that seeks to explain the responsiveness of flocks.

Although we observed variability in our computed values of \textit{m}* across different flocks, and variability was also observed in the estimated number of interacting neighbors for each flock in \cite{Ballerini08}, no correlation can be seen between these two values across flocks (Fig.~S9). This is not surprising, since any correlation would imply that all (or most) of the birds in a flock simultaneously change their number of interactions over time.

Although here we have focussed on the sensing strategy of interacting with \textit{m} nearest neighbors, our methods can also be applied to networks resulting from any sensing strategy. For example, our methods could be used to evaluate the robustness to noise of zonal sensing strategies like those used in \cite{Vicsek95, Hemelrijk11, Couzin05} as a function of a parameter such as zone size. Provided that a real or hypothesized sensing network can be constructed, its robustness can be calculated. However, care must be taken when comparing different strategies. For consistency, the weights in (\ref{eqn:dynamics}) should be scaled so that the sum of $a_{i,j}$ over all the neighbors of any individual (when neighbors are present) is equal to 1.

The nonlinear dependence on thickness observed in the random flocks suggests that a transition between ``2-d'' and ``3-d'' behavior takes place as thickness increases, with a flock behaving as fully 3-d when its thickness is above about 0.4. There appear to be aerodynamic reasons why starling flocks should be thin and sheet-like~\cite{Hemelrijk11}, and it is telling that the observed thicknesses lie near the transition point to fully three-dimensional behavior in terms of robustness. This suggests that groups with different characteristic thicknesses, such as  schools of fish, swarms of insects, and  herds of animals (with a thickness of zero), should interact with more (fewer) neighbors if they have a larger (smaller) thickness. Testing this hypothesis would provide important insight into the generality of this work for the analysis of animal groups. It should be noted, however, that factors that were not significant for starling flocks, such as flock width (Figs.~S4 and S5) or distribution (Figs.~S6 and S7) could play a larger role in 2-d groups.

More generally, our work demonstrates the significant role of who is interacting with whom in the ability of a network to efficiently manage uncertainty when seeking to maintain consensus. This suggests  possibilities for understanding and evaluating uncertainty management in other social and technological networks.    Our systems-theoretic approach to evaluating robustness to uncertainty in consensus can be applied to interaction networks in these other contexts; distinguishing interaction strategies that yield networks that optimize robustness can be useful both for  better understanding observed group behavior and, when control is available, for designing high performing groups.

\section*{Acknowledgments}
The authors thank three anonymous reviewers for helpful comments and suggestions.   A.C. and I.G. thank the Lewis-Sigler Institute for Integrative Genomics and the Physics Department at Princeton University for hospitality at the beginning of this work. Our collaboration was facilitated by the Initiative for the Theoretical Science at the Graduate Center of the City University of New York.


\newpage
\section*{Supporting Information}

\renewcommand{\thefigure}{S\arabic{figure}}
\setcounter{figure}{0}

\begin{flushleft}
\begin{figure}[!ht]
\centering
\includegraphics[width=\textwidth]{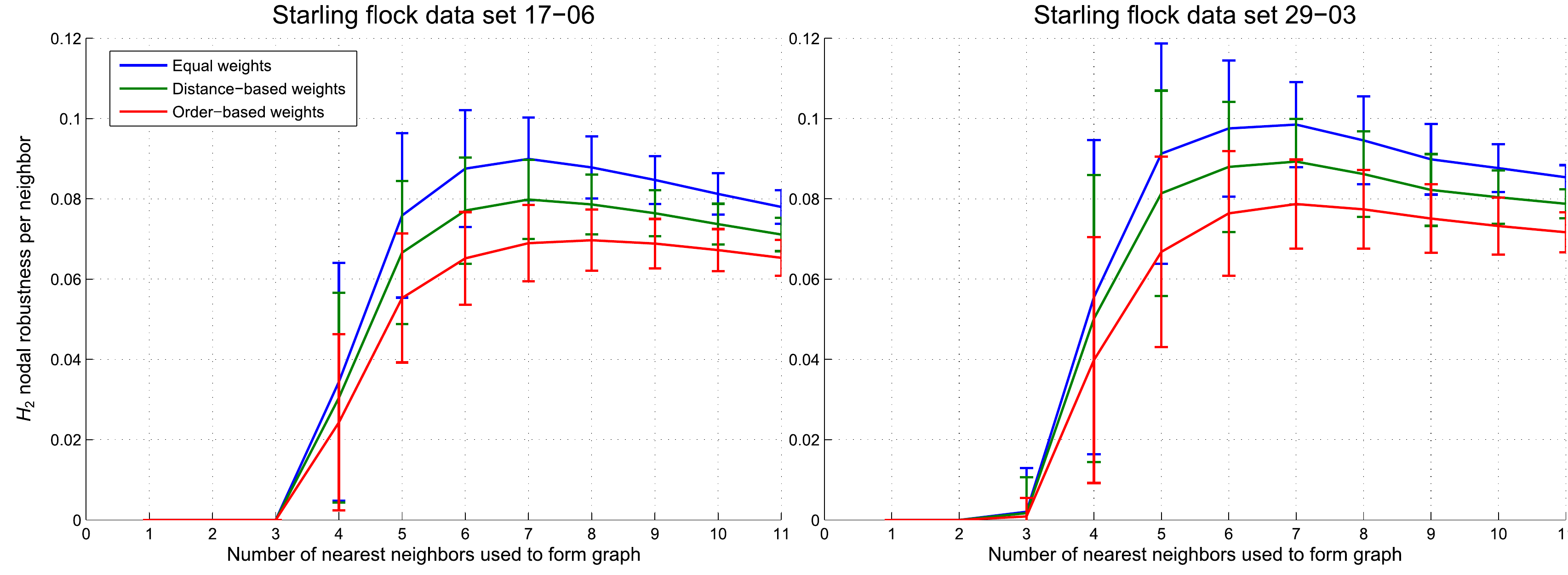}
\caption{\textbf{Average $H_2$ nodal robustness per neighbor as a function of the number of nearest neighbors (\textit{m}) used to form the graph for two different flocks, with the edge weights computed in three different ways.} Although equal edge weights were used throughout this paper, a plausible alternative is that greater weight is given to closer neighbors. In each case, the edge weights in the sensing graph are normalized so that the sum of all weights used by any individual bird is 1. The blue curves show results with equal edge weights (i.e.\ $a_{i,j} = \frac{1}{m}$ when an edge is present), as used in the rest of this paper. The green curves show results with edge weights inversely proportional to the distance between birds (i.e.\ $a_{i,j} \propto \frac{1}{d_{i,j}}$ when an edge is present, where $d_{i,j}$ is the distance between birds \textit{i} and \textit{j}). The red curves show results with edge weights decreasing linearly according to the ordering of the neighbors from closest to furthest, such that the $(m+1)^\text{st}$ neighbor has a weight of 0. Both additional weighting schemes decrease the importance of neighbors that are further away, although the order-based scheme is more ``radical'' since neighbors tend to be spaced closer than in a geometric progression. In every case, decreasing the weight given to further neighbors tends to decrease the overall robustness, but the location of the peak remains unchanged except for the order-based scheme in flock 17-06. These results suggest that a substantial variation in edge weights is required to move the peak of the robustness per neighbor curve, and furthermore that overall robustness is decreased by doing so.}
\label{fig:weights}
\end{figure}
\end{flushleft}

\begin{flushleft}
\begin{figure}[!ht]
\centering
\includegraphics[width=\textwidth]{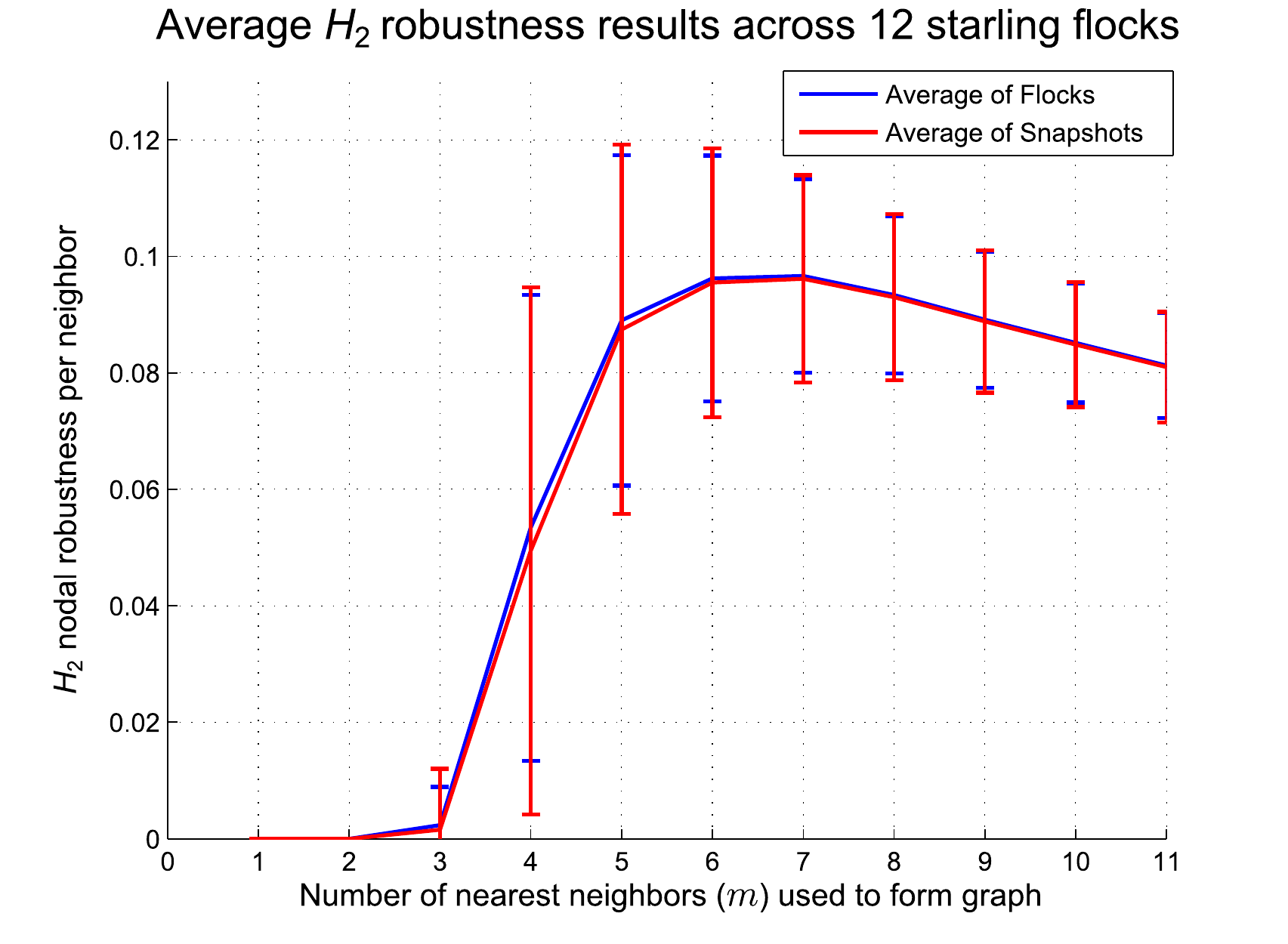}
\caption{\textbf{Average $H_2$ nodal robustness per neighbor as a function of the number of nearest neighbors (\textit{m}) used to form the graph, with the average taken in two different ways.} The blue curve shows the average of the twelve flock averages (as in Fig.\ 1 from the main text), while the red curve shows the average of the 394 snapshots taken across all flocks. In each case, the error bars show standard deviation. Since the results of the two averages match so closely while the number of snapshots taken of each flock varied between 16 and 80, this suggests that each snapshot may be taken to be an independent observation.}
\label{fig:snapshots}
\end{figure}
\end{flushleft}

\begin{flushleft}
\begin{figure}[!ht]
\centering
\includegraphics[width=\textwidth]{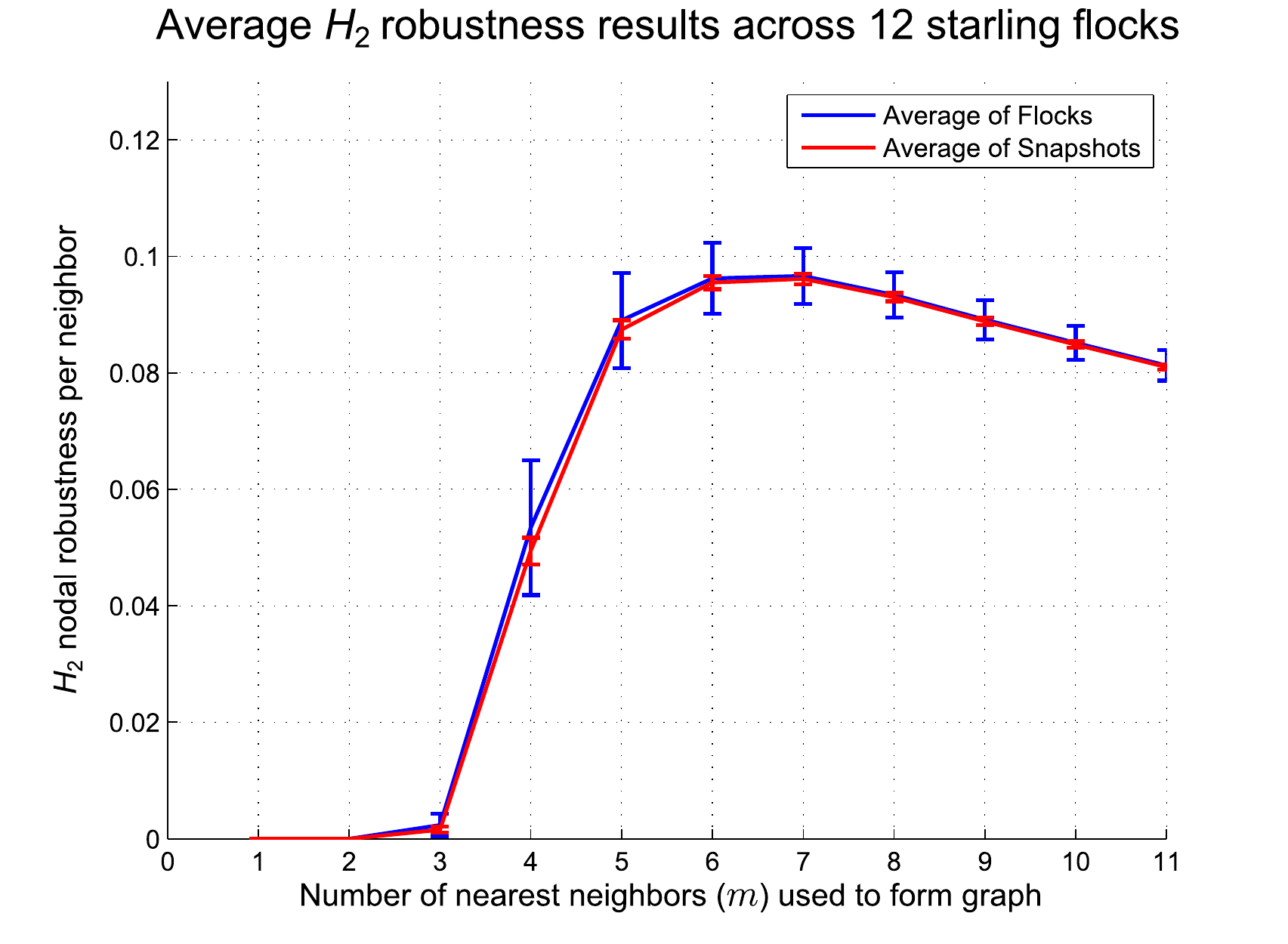}
\caption{\textbf{Average $H_2$ nodal robustness per neighbor with standard error as a function of the number of nearest neighbors (\textit{m}) used to form the graph, with the average taken in two different ways.} The blue curve shows the average of the twelve flock averages (as in Fig.\ 1 from the main text), while the red curve shows the average of the 394 snapshots taken across all flocks. In each case, the error bars show standard error. By treating each snapshot as independent (see Fig.~\ref{fig:snapshots}), the standard error is reduced and we can be more certain that the peak robustness per neighbor occurs at $m = 6$ or $m = 7$.}
\label{fig:error}
\end{figure}
\end{flushleft}

\begin{flushleft}
\begin{figure}[!ht]
\centering
\includegraphics[width=\textwidth]{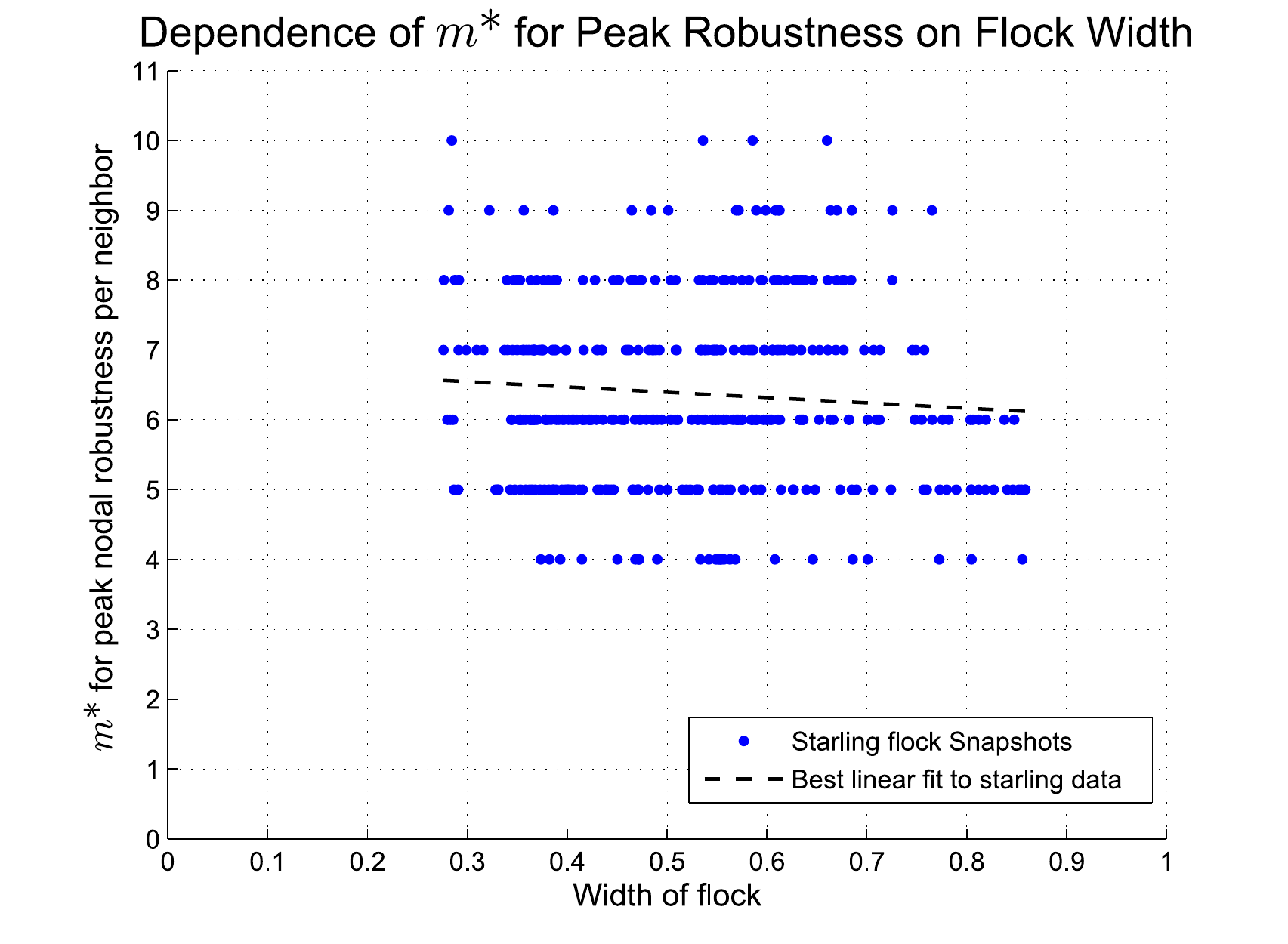}
\caption{\textbf{Dependence of the optimum number of neighbors (\textit{m}*) on the width of the flock.} Flock width is defined as the ratio of intermediate to largest dimension of an ellipsoid having the same principal moments of inertia as the flock. No significant dependence on width is observed, with the best linear fit having negligible slope and an $R^2$ value of 0.0064.}
\label{fig:mvw}
\end{figure}
\end{flushleft}

\begin{flushleft}
\begin{figure}[!ht]
\centering
\includegraphics[width=\textwidth]{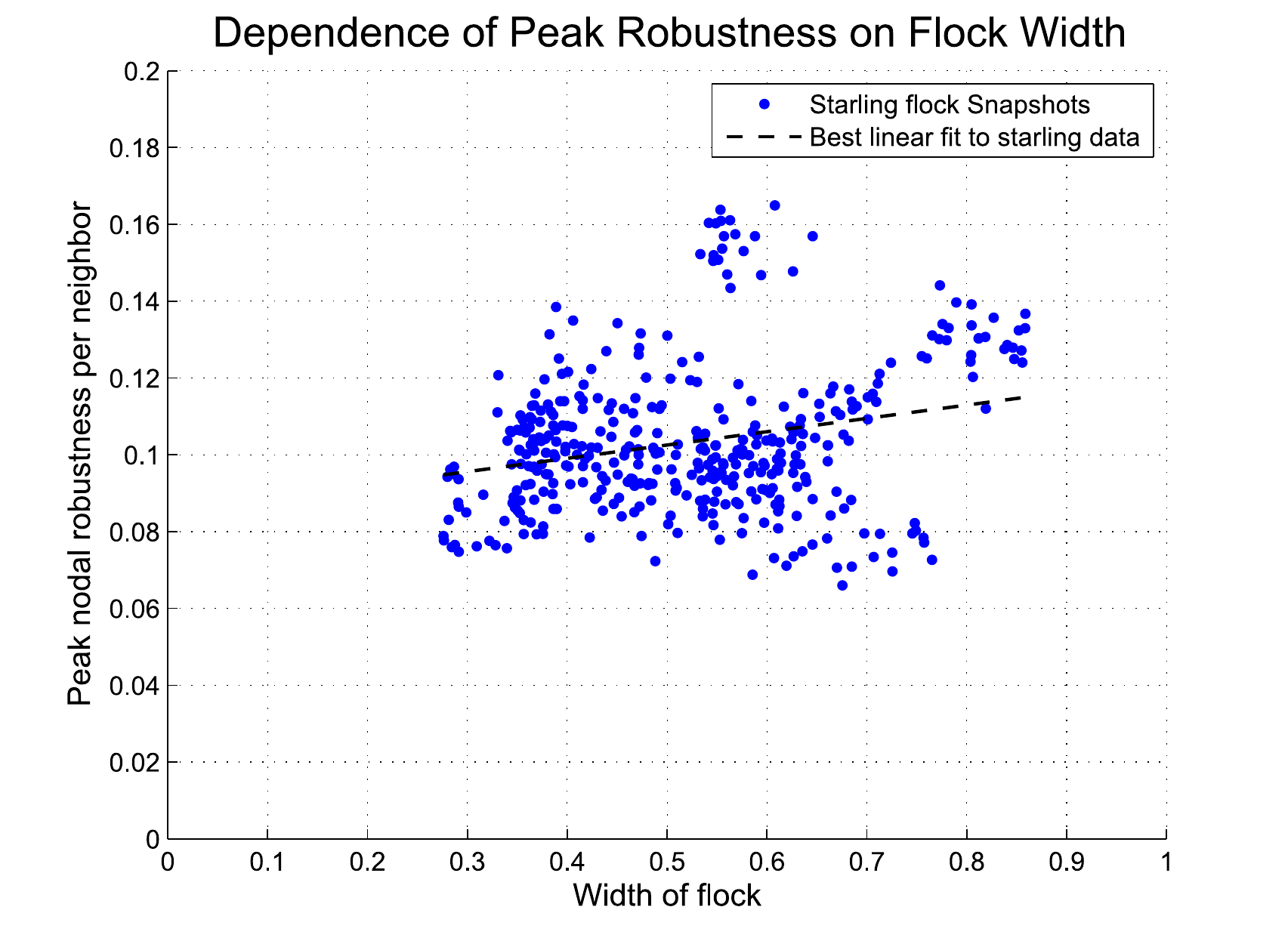}
\caption{\textbf{Dependence of the peak value of robustness per neighbor on the width of the flock.} Flock width is defined as the ratio of intermediate to largest dimension of an ellipsoid having the same principal moments of inertia as the flock. No significant dependence on width is observed, with the best linear fit having a slight positive slope and an $R^2$ value of 0.0635.}
\label{fig:pvw}
\end{figure}
\end{flushleft}

\begin{flushleft}
\begin{figure}[!ht]
\centering
\includegraphics[width=\textwidth]{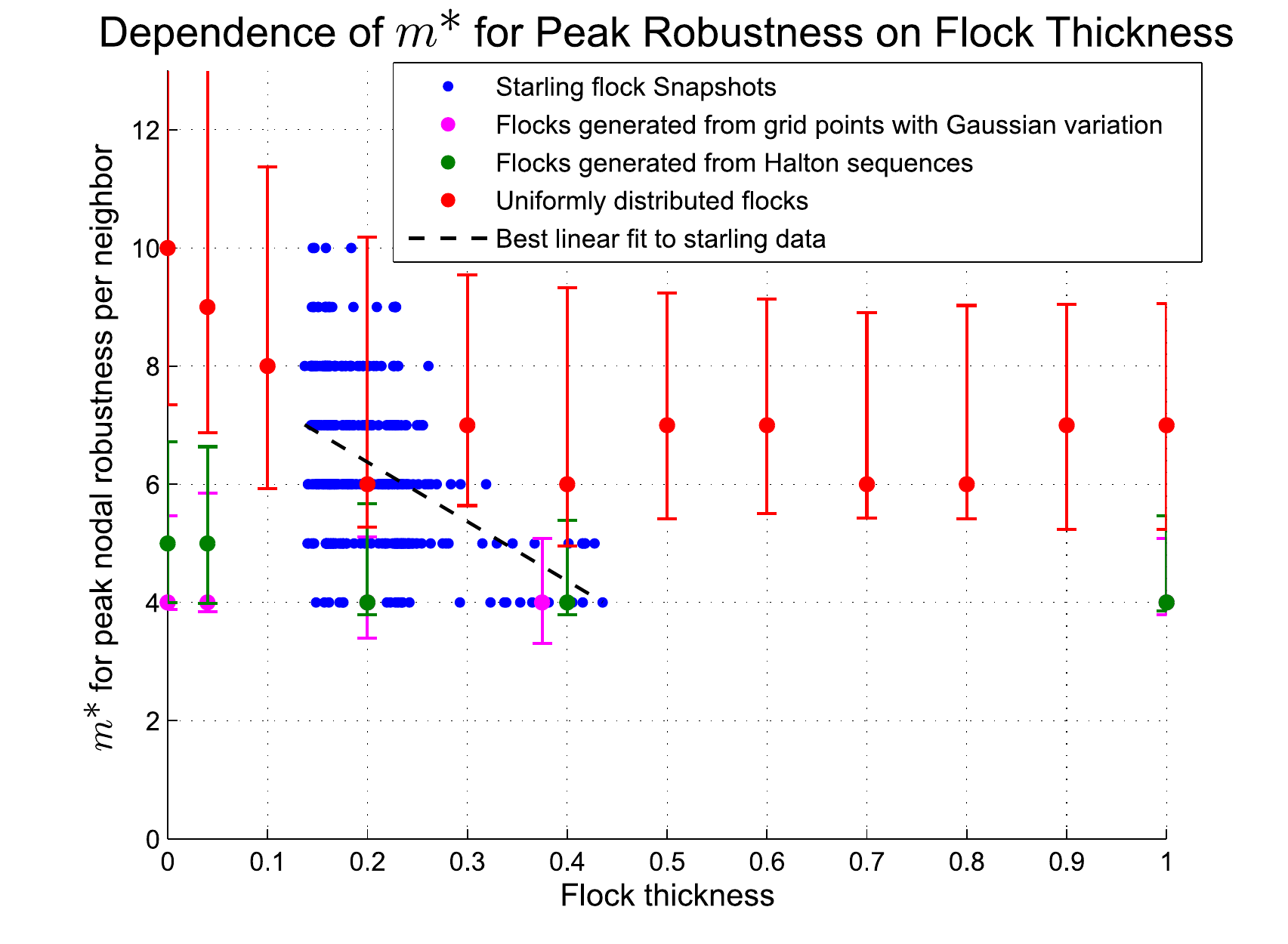}
\caption{\textbf{Dependence of the optimum number of neighbors (\textit{m}*) on the thickness of the flock.} In addition to starling data plotted in blue, results are shown from flocks randomly generated from three different distributions within a rectangular prism. These distributions are as follows: points arranged in a grid and then perturbed with Gaussian noise (in magenta), points generated from Halton sequences (in green) and points taken from a uniform distribution (in red). Each data point shown from the random flocks is the average result from generating 100 separate flocks, each containing approximately 1200 individuals. The error bars show the range of values for which the robustness per neighbor is within 90\% of the peak. As the random flocks become more ordered, the \textit{m}* values decrease and there is less of a dependence on thickness, with the most ordered flocks showing no thickness dependence. Compared to these three distributions, the starling flocks appear closest to uniform, with slightly more ``order.''}
\label{fig:mvt}
\end{figure}
\end{flushleft}

\begin{flushleft}
\begin{figure}[!ht]
\centering
\includegraphics[width=\textwidth]{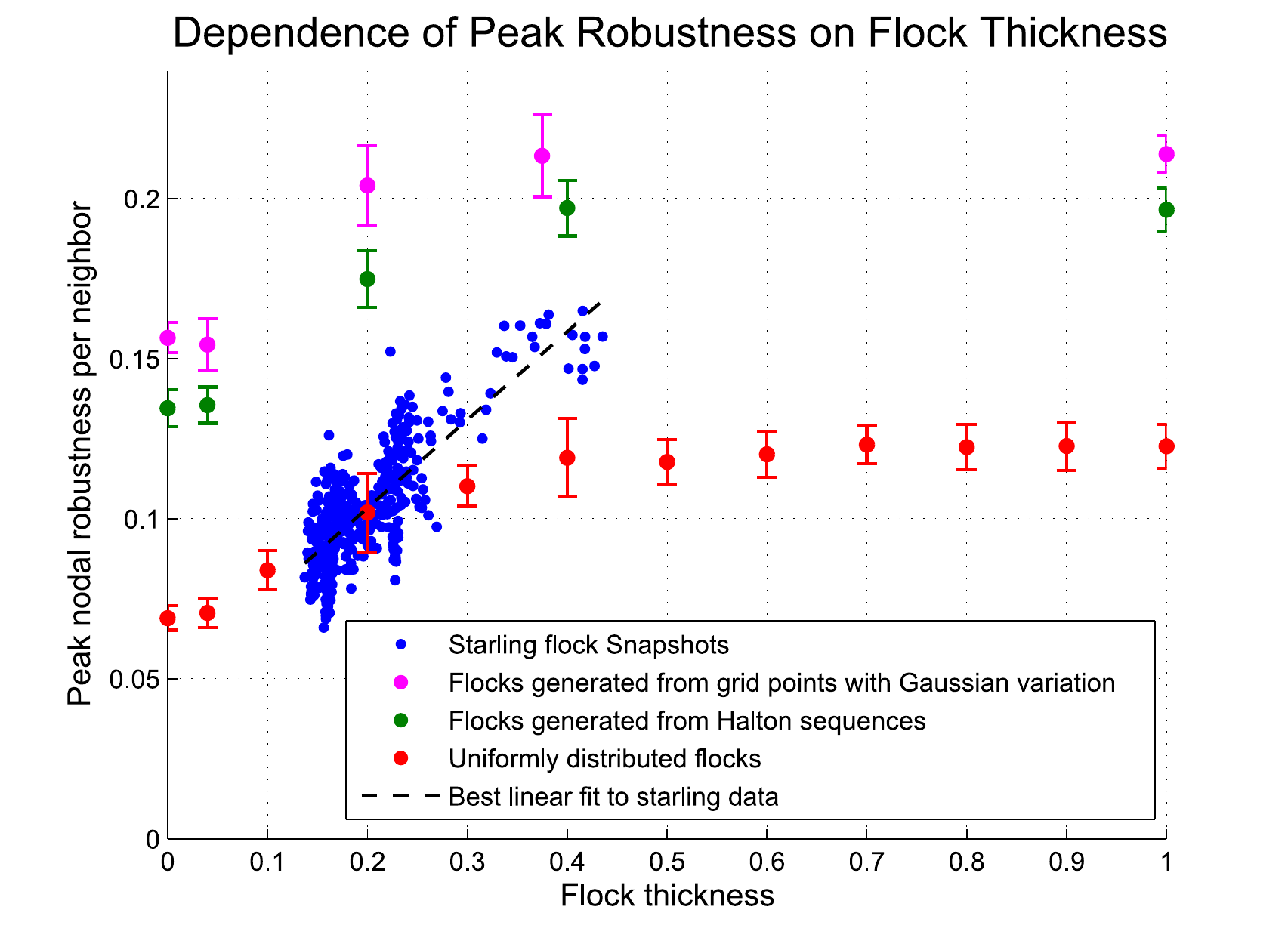}
\caption{\textbf{Dependence of the peak value of robustness per neighbor on the thickness of the flock.} In addition to starling data plotted in blue, results are shown from flocks randomly generated from three different distributions within a rectangular prism. These distributions are as follows: points arranged in a grid and then perturbed with Gaussian noise (in magenta), points generated from Halton sequences (in green) and points taken from a uniform distribution (in red). Each data point shown from the random flocks is the average result from generating 100 separate flocks, each containing approximately 1200 individuals. The error bars show standard deviation. As the random flocks become more ordered, the peak values increase but the same thickness trends are apparent, i.e., the curves all show a sigmoidal shape in the increase in peak value with increasing thickness. The starling flocks are close to the uniform flocks in most cases, with higher robustness values in other cases.}
\label{fig:pvt}
\end{figure}
\end{flushleft}

\begin{flushleft}
\begin{figure}[!ht]
\centering
\includegraphics[width=\textwidth]{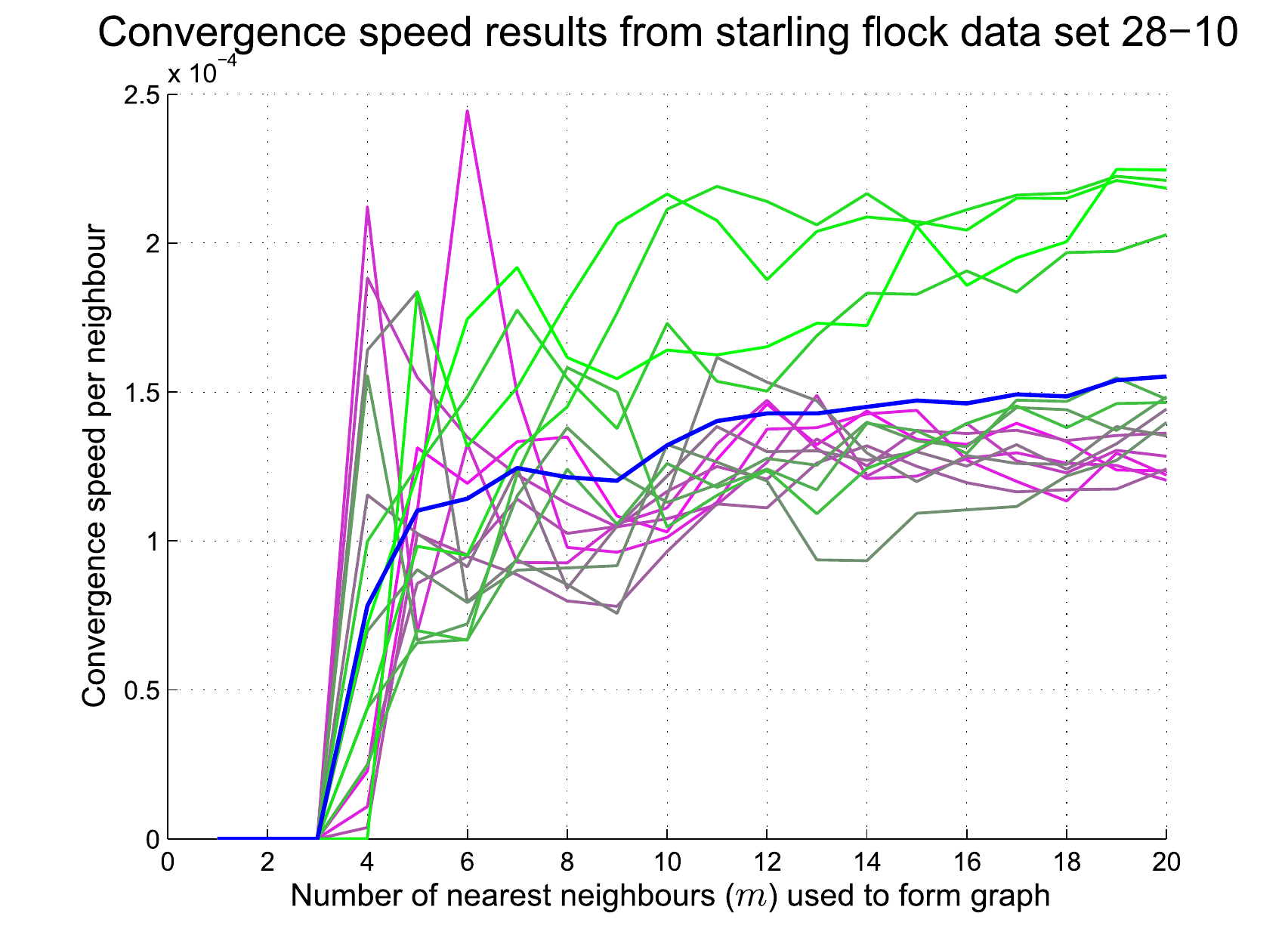}
\caption{\textbf{Speed of convergence to consensus (in the absence of noise) per neighbor as a function of the number of nearest neighbors (\textit{m}) used to form the graph, for one starling flock containing approximately 1300 birds.} Speed is computed as the real part of the second-smallest eigenvalue of the Laplacian matrix -- this is the exponential rate of convergence for the system in Eq.\  (2) of the main text. The units of speed are normalized with the fastest possible speed (in the case of every individual sensing every other individual) being 1. The thin lines show results for each snapshot, while the thick blue line shows the average over all snapshots. On average, speed per neighbor increases with \textit{m}, with no maximum observed for $m < 20$. In fact, when \textit{m} is equal to the number of birds in the flock, the speed per neighbor will be approximately $7.7 \times 10^{-4}$, significantly larger than the values for small \textit{m}.}
\label{fig:speed}
\end{figure}
\end{flushleft}

\begin{flushleft}
\begin{figure}[!ht]
\centering
\includegraphics[width=\textwidth]{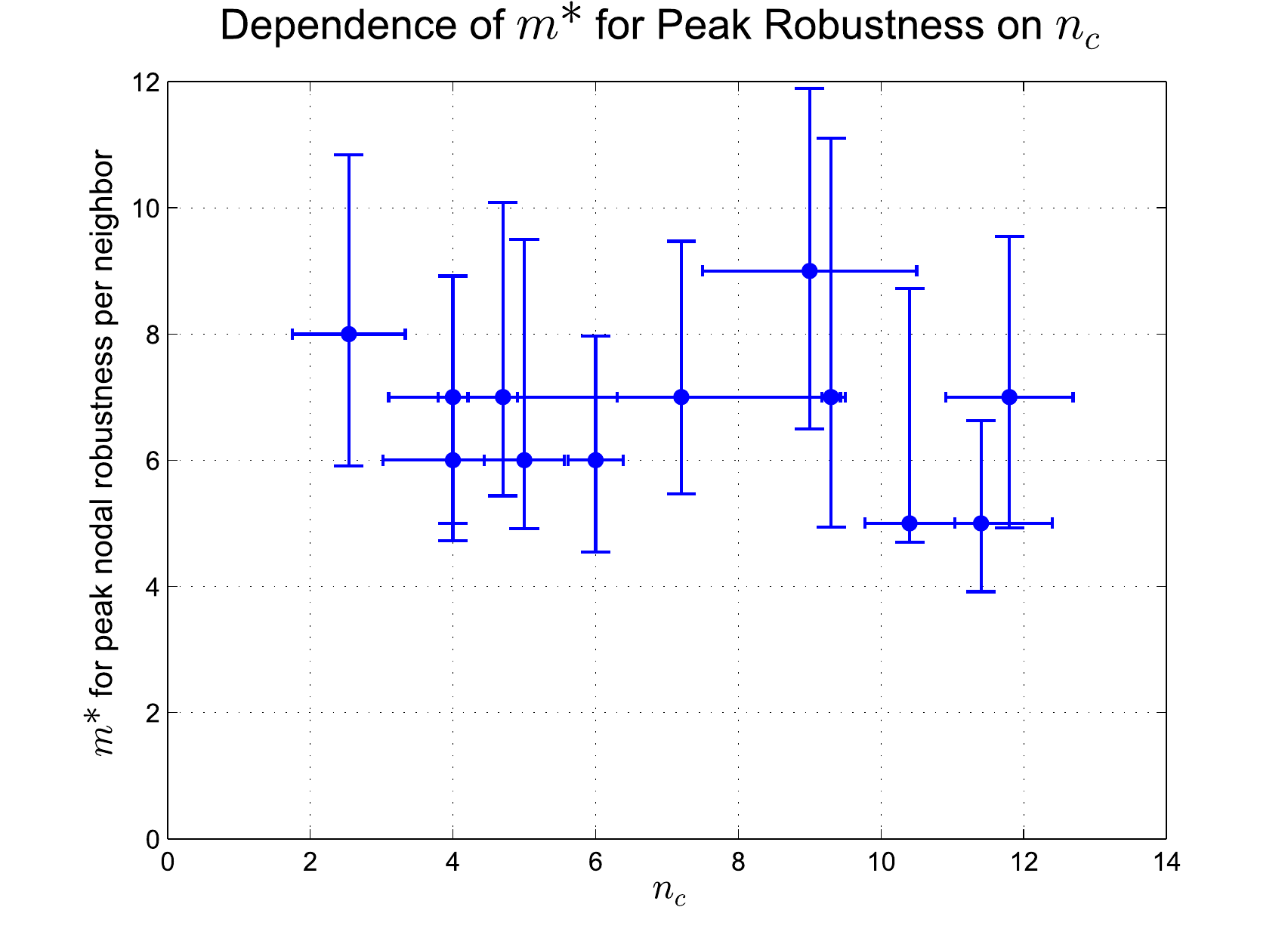}
\caption{\textbf{A comparison between the optimal number of neighbors (\textit{m}*) and the observed topological range ($n_c$) for each flock studied in this paper.} The vertical error bars show the range of values for which the robustness per neighbor is within 90\% of the peak and the horizontal error bars show the error in the estimates of $n_c$. No significant correlation is observed between these two measures, with a correlation coefficient of approximately $-0.24$ and a p-value of approximately 0.46. This is not surprising since it seems unlikely that individual starlings would interact with more or fewer neighbors based on the thickness of the flock (or any other bulk flock parameter). In addition, this suggests that the two analyses are independent and there is no underlying mathematical reason why \textit{m}* should be so close to $n_c$.}
\label{fig:correlation}
\end{figure}
\end{flushleft}


\end{document}